\documentclass{appolb}
\usepackage{epsfig}

\renewcommand{\vec}[1]{\mbox{\boldmath $#1$}}  
\newcommand{\tdm}[1]{\mbox{\boldmath $#1$}}

\newcommand{\be}{\begin{equation}}
\newcommand{\ee}{\end{equation}}
\newcommand{\ba}{\begin{eqnarray}}
\newcommand{\ea}{\end{eqnarray}}
\newcommand{\bc}{\begin{center}}
\newcommand{\ec}{\end{center}}
\newcommand{\bfig}{\begin{figure}}
\newcommand{\efig}{\end{figure}}

\begin{document}

\title{Diffractive vector mesons at 
large momentum transfer\\from the BFKL equation\thanks{Presented by RE at 
X International Workshop on Deep Inelastic Scattering (DIS2002)
Cracow, May 2002.
}}
\author{R.~Enberg$^{a}$, L.~Motyka$^{a,c}$ and
G.~Poludniowski$^{b}$
\address{
$^{a}$ High Energy Physics, Uppsala University, Sweden\\
$^{b}$ Department of Physics and Astronomy, University of Manchester, UK \\
$^{c}$ Institute of Physics, Jagellonian University, Krak\'{o}w, Poland \\ 
}
}
\maketitle
\begin{abstract}
Diffractive vector meson photoproduction accompanied by proton
dissociation is studied for large momentum transfer.
The process is described by the non-forward BFKL equation, for which an 
analytical solution is found for all conformal spins,
giving the scattering amplitude. Results are compared to HERA
data on $\rho$ production.
\end{abstract}

Diffractive production of vector mesons in $\gamma p$ collisions at large
momentum transfer, $\gamma p \to VX$, is an experimentally clean process.
The signal consists of an isolated vector meson with large transverse 
momentum, separated from the remnant of the incoming proton by a large rapidity
gap. There are recent measurements of cross-sections and helicity amplitudes
for this process \cite{ZEUS,H1}.

The large momentum transfer involved makes it possible to describe the colour
singlet exchange in terms of perturbative QCD. This is in contrast to vector
meson production in diffractive processes with small momentum transfer, where
the sensitivity to the infrared region is larger.
The perturbative QCD description of hard colour singlet exchange across a large
rapidity interval relies on the BFKL equation \cite{BFKL1,BFKL2}, 
which resums leading powers of
the rapidity $y$ to all orders in the perturbative expansion of the amplitude.
The colour singlet system, or pomeron, is here a composite system of two
reggeized gluons.

Let us mention that this process has been studied before; for heavy mesons
in BFKL \cite{HVMold} using the Mueller-Tang \cite{MT} approximation, and
for light mesons at the Born level \cite{IKSS}. The data on the cross-sections 
can be fitted with a BFKL calculation, but not by the fixed order formulae.
The understanding of the helicity structure remains a challenge.

We have recently studied the production of heavy mesons in a BFKL 
framework \cite{HVM}, so here we concentrate on the case of light mesons.


At large momentum transfer the pomeron couples predominantly to a single
parton, (see fig.\ \ref{Feyn}), 
which means that the cross-section may be factorized into a
convolution of the partonic cross-section with the parton distribution
functions. We therefore calculate the amplitude for
$\gamma q\to V q$ (since $\gamma g\to V g$ differs only by a colour factor).
 
In the BFKL framework the scattering amplitude is calculated as the convolution
of three factors; schematically
$\mathcal{A} = 
\Phi^{\gamma\to V}\otimes K_{\mathrm{BFKL}} \otimes \Phi^{q\to q},
$
where $\Phi$ are the impact factors describing the coupling of the 
pomeron to the indicated vertices, and $K_{\mathrm{BFKL}}$ is the BFKL kernel
describing the evolution of the gluon ladder.

The non-forward BFKL equation has a solution due to Lipatov \cite{BFKL2}:
\be
 \mathcal{A}   = \frac{1}{(2\pi)^6}  \sum_{n} 
 \int d\nu  \,
\frac{{\nu^2 + \frac{{n}^2}{4}}}
{[\nu^2 + (\frac{{n}-1}{2})^2]
[\nu^2+(\frac{{n}+1}{2})^2]} 
\;e^{\omega_n(\nu) y}
{{I^1 _{{n},\nu}}
(\tdm k,\tdm q)\, {I^{2\,^\star}_{{n},\nu}} (\tdm k',\tdm q)}.
\label{Solution}
\ee
This represents an expansion of the amplitude in the complete basis of
eigenfunctions $E_{n,v}$ of the BFKL kernel. 
The functions ${I^{1,2}_{{n},\nu}}$ are projections of the impact factors 
$\Phi^{1,2}(\tdm k,\tdm q)$ on these eigenfunctions, see
\cite{BFKL2,HVM} for details.

The integer $n$ in (\ref{Solution}) is known as the {\em conformal spin}.
The terms in the sum with non-zero $n$ are exponentially suppressed by the
factor $e^{\omega_n(\nu) y}$, with $\omega_n(\nu) <0$ for $n\neq0$, 
and so the amplitude 
is usually approximated by the leading $n=0$ term (the
Mueller-Tang approximation \cite{MT}). This
approximation, however, is only good for very large rapidities $y$. 
For moderate $y$ the
higher $n$ terms can still be important, as was found in \cite{MMR,EIM}. We
therefore calculate the amplitudes including all $n$.

 The quark impact factor is given in ref.\ \cite{MMR} for all
conformal spins, and we have to compute the vector meson impact factors. This is
done separately for heavy \cite{HVM} and light \cite{LVM}
mesons, using different approximations for the vector meson wave functions.
In the heavy meson case, we used the non-relativistic approximation,
where the constituent quarks are assumed to each
carry half of the meson momentum. Our results can be found in \cite{HVM}.

Ivanov {\it et al.\/} \cite{IKSS}
give the helicity amplitudes for light meson production,
$M_{++}$, $M_{+0}$ and $M_{+-}$, where the first index corresponds to
the polarization of the incoming photon and the second to the vector meson.
These are referred to as the no-flip, single-flip and double-flip amplitudes, 
respectively. 

Their calculation assumes
Born-level two-gluon exchange, and they use a relativistic approximation for
the vector meson wave functions, with massless quarks.
Note that the longitudinal and transverse degrees of freedom are factorized.
For instance, taking $\vec r$ to be the transverse separation of the quarks in 
the vector meson, and $u$ to be the lightcone momentum fraction of the quark,
they give the single-flip amplitude as

\be
M_{+0}= 
\; 
\int 
{\frac{d^{\,2}\vec k}{\vec k^2(\vec k-\vec q)^2}} 
\left[C \, \alpha_s^2 \; \int
\frac{d^2\vec r\,du}{4\pi}\; 
{f^{\mathrm{dipole}}}\;
\frac{\vec r\cdot\vec e^+}{r^2}\;
\frac{f_\rho}{2} \;  (1-2u) 
{\phi_{\|}(u)} \right],
\label{M+0}
\ee 
where $\phi_{\|}(u)$ is the twist-2 vector meson wave function.
Expressed in the impact factor picture, the factor in brackets is just the
product of the impact factors $\Phi^{q\to q}$ and $\Phi_{+0}^{\gamma\to V}$.
Thus, we calculate $I^{\gamma\to V}_{n,\nu}$ by projecting
these expressions onto the BFKL eigenfunctions, and insert the result into 
(\ref{Solution}).

The projection of the impact factor $I^{\gamma\to V}_{n,\nu}$ is then
proportional to \cite{LVM}
\ba
\mathcal I &=& \sin \pi \left( 1/2+\beta+\widetilde\mu\right ) \;
         B_+( \alpha, q^*, \xi^* ) \;
         \widetilde B_+( \beta, q, \xi) 
         \nonumber\\
&-& \sin \pi \left( 1/2+\beta-\widetilde\mu\right ) \;
         B_-( \alpha, q^*, \xi^*) \;
         \widetilde B_-( \beta, q, \xi ) 
\ea
for even $n$ and zero for odd $n$.
Here $\mu=n/2-i\nu, \widetilde\mu=-n/2-i\nu$, 
and we introduce the conformal blocks $B_\pm$
\ba
B_\pm( \alpha, q, \xi ) &=& 
\left(   \frac{2i}{k}  \right) ^{\frac{3}{2}+\alpha}
\left(   \frac{iq}{4k}  \right) ^{\pm\mu}
         \frac{\Gamma(3/2+\alpha \pm\mu)}{\Gamma(1\pm\mu)}
         \nonumber\\
&\times& _2F_1\left(
         3/2 + \alpha   \pm\mu \, , \;
         1/2 \pm\mu\,;\;
         1\pm2\mu\,;\;
         2/(1+\xi)      \right),
\ea
and $\widetilde B_\pm$ obtained by $\mu \to \widetilde\mu$. The constants
$\alpha$ and $\beta$ take different values for different helicity amplitudes.

\bfig
\bc
\epsfig{file=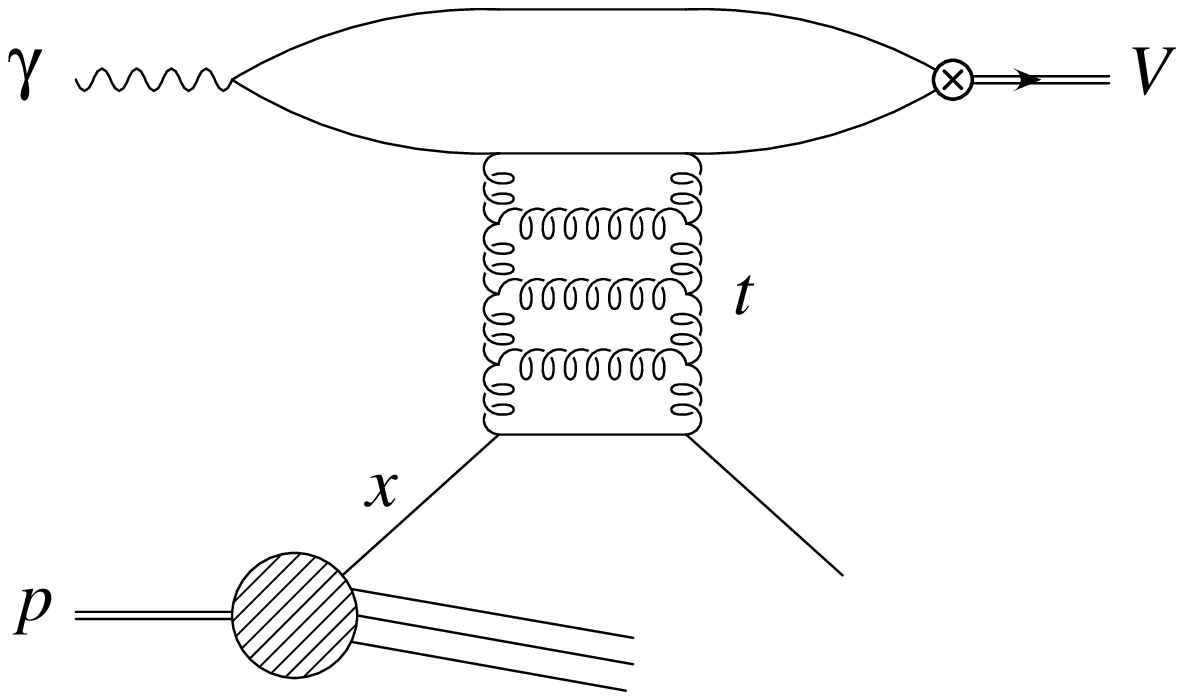, width=0.49\columnwidth,
        bbllx=173, bblly=286,bburx=510, bbury=517}
\epsfig{file=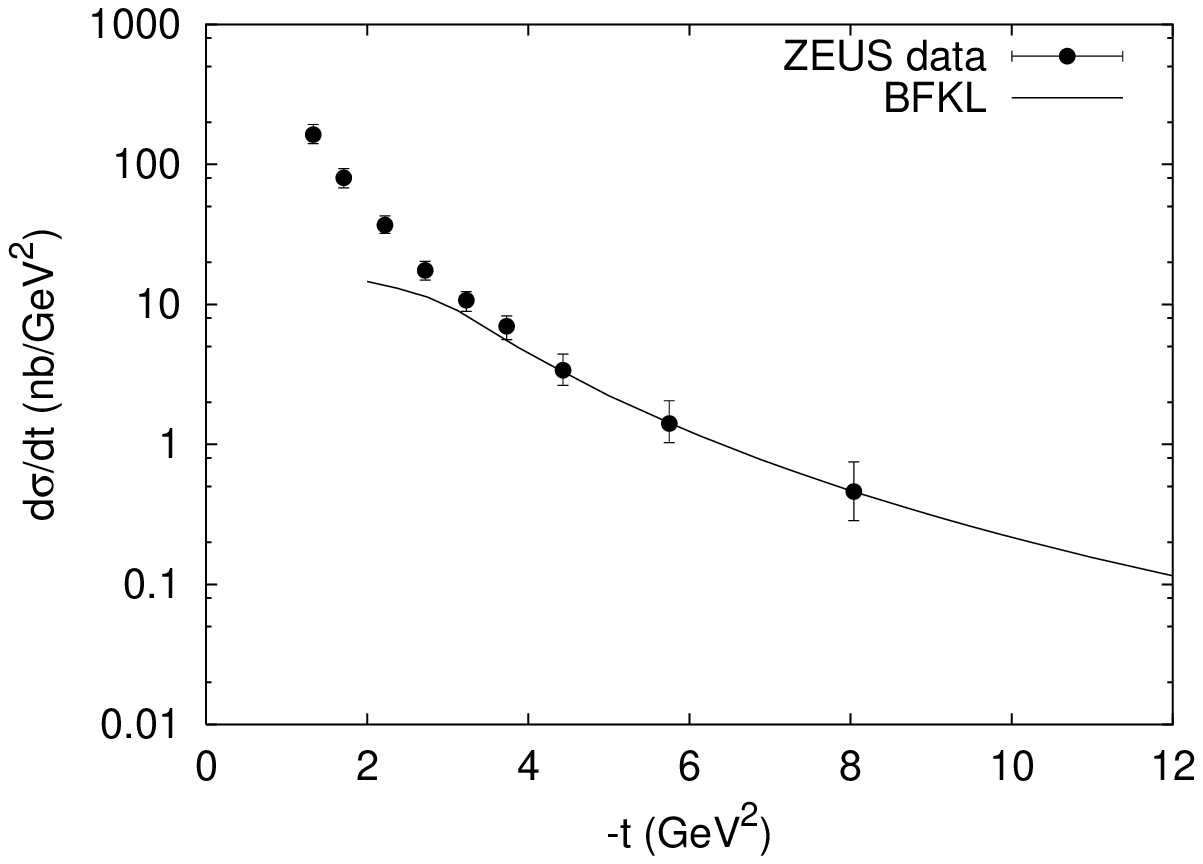, width=0.5\columnwidth}
\ec
\caption{Feynman graph and differential cross-section for 
the process $\gamma p\to\rho X$.}
\label{Feyn}
\label{rhosigma}
\efig
We will now go on to evaluate the obtained expressions,
concentrating on the case of light mesons and referring to \cite{HVM} 
for our results on heavy mesons. 

ZEUS have measured the differential cross-section $d\sigma/dt$ for 
$\rho$ and $\phi$ mesons for momentum transfers up to 8 GeV$^2$ \cite{ZEUS}. 
In fig. \ref{rhosigma} we show a comparison of the data together with our
calculation, including all conformal spins. Here we have chosen a fixed value 
of $\alpha_s=0.38$ in the
prefactor (see (\ref{M+0})) and $\alpha_s=0.24$ in the eigenfunctions
$\omega_n(\nu)$, defining the pomeron intercept. These choices of different 
$\alpha_s$ reflect the impact of non-leading corrections to the BFKL intercept.
The rapidity is defined as
$y=\ln (\hat s/m_\rho^2)$. Note that the turnover at $|t| \sim 2$ GeV$^2$ is 
due to the too restrictive infrared cut-off $u_{min}= -m_\rho^2/t$ 
(taken from \cite{IKSS}) in the integration over $u$. Such a cut-off
in $u$ was required to regulate an unphysical divergence.

In addition, ZEUS have also measured the spin density matrix elements 
$r_{ij}^{04}$ for the process. These parametrize the decay angular distributions
of the mesons, and can be related to the helicity amplitudes $M_{ij}$ (see e.g.\
\cite{ZEUS}). Note, however, that both the no-flip and double-flip amplitudes
processes lead to the same polarization states of the vector meson, and
therefore they cannot be distinguished in unpolarized experiment. 
The $r_{ij}^{04}$ contain interferences,
though, so it can be inferred from the data that all three helicity amplitudes
are non-zero for $\rho$ and $\phi$, while for $J/\Psi$, only the no-flip
amplitude seems to be non-zero.

\bfig
\bc
\epsfig{file=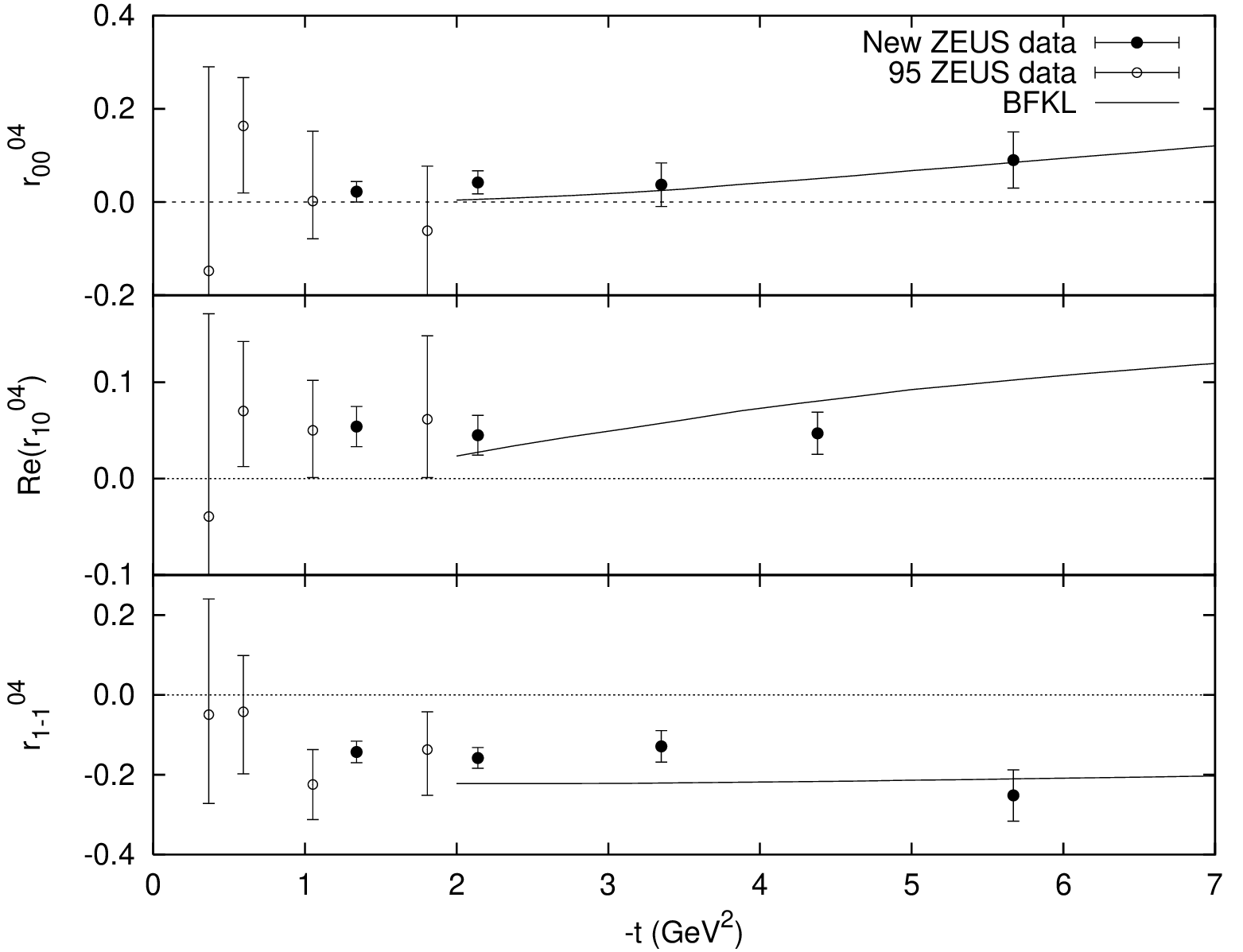, width=0.8\columnwidth,
        bbllx=50, bblly=115,bburx=712, bbury=622}
\ec
\caption{Spin density matrix elements for $\rho$ meson production.}
\label{r}
\efig
In fig.\ \ref{r}, we show our BFKL predictions for $r_{ij}^{04}$ compared
to the ZEUS data. The qualitative features are well reproduced. The shapes of
the curves depend somewhat on the choices of pomeron intercept and definition of
the rapidity, but we find that 
({\em i}\/) the $(+-)$ component of the amplitude dominates and 
({\em ii}\/) the $(++)$ component is negative. The BFKL evolution thus
changes the features from the fixed-order Born level results of \cite{IKSS}.

These results should be interpreted with some care, however, because of
some uncertainties. We have not included
the chiral-odd components of the photon wave function of \cite{IKSS},
but believe that these may be small \cite{LVM}.
Also, the endpoints in the $u$-integration have to be treated carefully and may
change the results. Furthermore, the approximations used, with massless quarks 
and a factorized meson wave function, have to be understood.

In conclusion, we have studied the process $\gamma p \to V X$
in a BFKL framework, obtaining exact solutions of the BFKL
equation. 
Comparing the calculations to ZEUS data shows good agreement with both the
differential cross-section and the spin density matrix elements for diffractive
$\rho$ production.

We thank Jeff Forshaw for useful discussions. 
This study was supported in part by the Swedish Natural Science
Research Council and by the Polish Committee for Scientific
Research (KBN) grant no. 5P03B~14420.


\end{document}